\documentclass{ws-procs975x65}

\usepackage{times}
\usepackage{endnotes}

\usepackage{graphicx}
\usepackage{color}
\usepackage{dcolumn}
\usepackage{bm}

\def\comment#1{}

\def\secton#1{\section{#1}}

\begin{document}

\title{
  Architecture-Dependent Execution Time of Shor's Algorithm
}

\author{Rodney Van Meter$^{1}$\thanks{E-mail address:
    rdv@tera.ics.keio.ac.jp.}, Kohei M. Itoh$^{1}$ and
  Thaddeus D. Ladd$^{2}$ \\
$^{1}$Graduate School of Science and Technology, Keio
  University and CREST-JST\\
3-14-1 Hiyoshi, Kohoku-ku, Yokohama-shi, Kanagawa 223-8522, Japan \\
$^{2}$Edward L. Ginzton Laboratory \\
Stanford University, Stanford, CA, 94305-4085, USA}





\maketitle

\abstracts{We show how the execution time of algorithms on quantum
computers depends on the architecture of the quantum computer, the
choice of algorithms (including subroutines such as arithmetic), and
the ``clock speed'' of the quantum computer.  The primary
architectural features of interest are the ability to execute multiple
gates concurrently, the number of application-level qubits available,
and the interconnection network of qubits.  We analyze Shor's
algorithm for factoring large numbers in this context.  Our results
show that, if arbitrary interconnection of qubits is possible, a
machine with an application-level clock speed of as low as one-third
of a (possibly encoded) gate per second could factor a 576-bit number
in under one month, potentially outperforming a large network of
classical computers.  For nearest-neighbor-only architectures, a clock
speed of around twenty-seven gates per second is required.}


\bibliographystyle{apsrev}

\secton{Introduction}

Quantum computers are currently being designed that will take
advantage of quantum mechanical effects to perform certain
computations much faster than can be achieved using current
(``classical'') computers~\cite{nielsen-chuang:qci}.  Many
technological approaches have been proposed, some of which are being
investigated experimentally.  DiVincenzo proposed five criteria which
must be met by any useful quantum computing
technology~\cite{divincenzo95:qc}.  In addition to these criteria, a
useful quantum computing technology must also support a quantum
computer {\em system architecture} which can run one or more quantum
algorithms in a usefully short time.  This observation subsumes into
one requirement several issues which, while not strictly necessary to
build a quantum computer, will have a strong impact on the possibility
of engineering a practical system.  These include the importance of
gate ``clock'' speed, support for concurrent gate operations, the
total number of application-level qubits supportable, and the
complexities of the qubit interconnect
network~\cite{van-meter:qarch-impli}.

\begin{figure}
\centerline{\hbox{
\includegraphics[width=8.6cm]{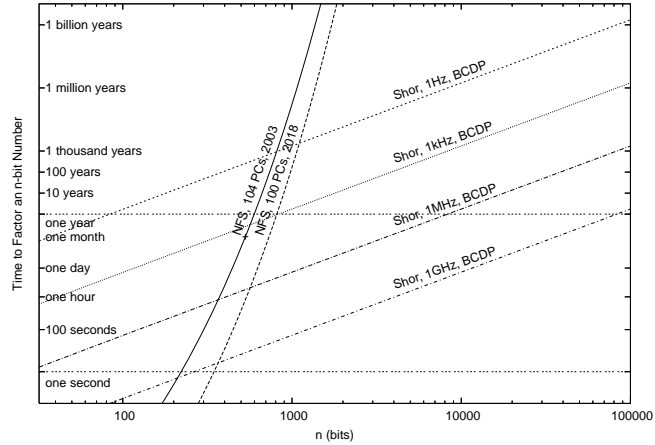}}}
\caption{Scaling of number field sieve (NFS) on classical computers
  and Shor's algorithm for factoring on a quantum computer, using
  BCDP modular exponentiation with various clock rates.  Both
  horizontal and vertical axes are log scale.  The horizontal axis is
  the size of the number being factored.}
\label{fig:scaling-bcdp}
\end{figure}

This paper discusses the impact of these architectural elements on
algorithm execution time using the example of Shor's algorithm for
factoring large numbers~\cite{shor:factor}.  Shor's algorithm ignited
much of the current interest in quantum computing because of the
improvement in computational class it appears to offer on this
important problem.  Using Shor's algorithm, a quantum computer can
solve the problem in polynomial time, for a superpolynomial speed-up.
Shor's algorithm is theoretically important, well defined, and
utilizes building blocks (arithmetic, the quantum Fourier transform)
with broad applicability, making it ideal for our analysis.

On a classical computer, or a collection thereof, the time and
computing resources to factor a large number, using the fastest known
algorithm, scale superpolynomially in the length of the number (in
decimal digits or bits).  This algorithm is the generalized number
field sieve (NFS)~\cite{knuth:v2-3rd}.  Its asymptotic computational
complexity on large numbers is
\begin{equation}
\label{eq:nfs}
O(e^{(nk \log^2 n)^{1/3}})
\end{equation}
where $n$ is the length of the number, in bits, and $k =
\frac{64}{9}\log 2$.  The comparable computational complexity to
factor a number $N$ using Shor's algorithm is dominated by the time to
exponentiate a randomly chosen number $x$, modulo $N$, for a
superposition of all possible exponents.  Therefore, efficient
arithmetic algorithms for calculating modular exponentiation in the
quantum domain are critical.

Very often clock speed and other architectural features are ignored as
issues in quantum computing devices, assuming that the superpolynomial
speed-up will dominate, making the algorithm practical on any
experimentally realizable quantum computer.  Shor's algorithm runs in
polynomial time, but the details of the polynomial matter: what degree
is the polynomial, and what are the constant factors?

An immediate comparison of the execution time to factor a number on
classical and quantum computers is shown in
Figure~\ref{fig:scaling-bcdp}.  The performance of Shor's algorithm on
a quantum computer using the Beckman-Chari-Devabhaktuni-Preskill
(BCDP) modular exponentiation
algorithm~\cite{beckman96:eff-net-quant-fact} is compared to classical
computers running the general Number Field Sieve (NFS).  The steep
curves are for NFS on a set of classical computers.  The left curve is
extrapolated performance based on a previous world record, factoring a
530-bit number in one month, established using 104 PCs and
workstations made in 2003~\cite{rsa04-factoring}.  The right curve is
speculative performance using 1,000 times as much computing power.
This could be 100,000 PCs in 2003, or, based on Moore's law, 100 PCs
in 2018.  From these curves it is easy to see that Moore's law has
only a modest effect on our ability to factor large numbers.  The
shallower curves on the figure are predictions of the performance of a
quantum computer running Shor's algorithm, using the BCDP modular
exponentiation routine, which uses $5n$ qubits to factor an $n$-bit
number, requiring $\sim 54n^3$ gate times to run the algorithm on
large numbers.  The four curves are for different clock rates from
1~Hz to 1~GHz.  The performance scales linearly with clock speed.
Factoring a 576-bit number in one month of calendar time requires a
clock rate of 4~kHz.  A 1~MHz clock will solve the problem in about
three hours.  If the clock rate is only 1~Hz, the same factoring
problem will take more than three hundred years.

The performance of the BCDP modular exponentiation algorithm is almost
independent of architecture.  However, the performance of most
polynomial-time algorithms varies noticeably depending on the system
architecture~\cite{kunihiro05:_exact_ieice,fowler04:_shor_implem}.
The main objective of this paper is to show how we can improve the
execution time shown in Figure~\ref{fig:scaling-bcdp} by understanding
the relationship of architecture and algorithm.

\secton{Results}

We have analyzed two separate architectures, still technology
independent but with some important features that help us understand
performance. The AC ({\em abstract concurrent}) architecture is our
abstract model, akin to what is commonly used when drawing quantum
circuits.  It supports arbitrary concurrency and gate operands any
distance apart without penalty.  The second architecture, NTC ({\em
neighbor-only, two-qubit gate, concurrent}) , assumes the qubits are
laid out in a one-dimensional line, and only neighboring qubits can
interact.  This is a reasonable description of several important
experimental approaches, including a one-dimensional chain of quantum
dots~\cite{loss:qdot-comp}, the original Kane
proposal~\cite{kane:nature-si-qc}, and the all-silicon NMR
device~\cite{ladd:si-nmr-qc}.

Above the architecture resides the choice of algorithm, especially for
basic arithmetic operations.  The computational complexity of an
algorithm can be calculated for total cost, or for {\em latency} or
{\em circuit depth}, if the dependencies of variables allow multiple
parts of a computation to be conducted concurrently.  Fundamentally,
the computational complexity of quantum modular exponentiation is
$O(n^3)$~\cite{vedral:quant-arith,beckman96:eff-net-quant-fact}, that
is, the execution cost grows as the cube of the number of qubits.  It
consists of $2n$ modular multiplications of $n$-bit numbers, each of
which consists of $O(n)$ additions, each of which requires $O(n)$
operations.  However, $O(n^3)$ {\em operations} do not necessarily
require $O(n^3)$ {\em time steps}; the {\em circuit depth} can be made
shallower than $O(n^3)$ by performing portions of the calculation
concurrently.

On an abstract machine, we can reduce the running time of each of the
three layers (addition, multiplication, exponentiation) to $O(\log n)$
time steps by running some of the gates in parallel, giving a total
running time of $O(\log^3 n)$. This requires $O(n^3)$ qubits and the
ability to execute an arbitrary number of gates on separate qubits.
Such large numbers of qubits are not expected to be practical for the
foreseeable future, so interesting engineering lies in optimizing for
a given set of architectural constraints.

Addition forms the basis of multiplication, and hence of
exponentiation.  Classically, many forms of adders have been used in
computer hardware~\cite{ercegovac-lang:dig-arith}.  The most basic
type of adder, variants of which are used in both VBE and BCDP (as
well as our algorithm {\bf F}, below), is the {\em carry-ripple}
adder, in which the carry portion of the addition is done linearly
from the low-order bits to the high-order.  This form of adder is
$O(n)$ in both circuit depth and complexity; it is the only efficient
type for NTC linear architectures, in which the time to propagate
the low-order carry is inherently constrained to $O(n)$.  When
long-distance gates are available, as in AC architectures, the use of
faster adders such as conditional-sum, carry-lookahead, or carry-save
adders can result in $O(\log n)$ latency, though the complexity
remains
$O(n)$~\cite{van-meter04:fast-modexp,draper04:quant-carry-lookahead,gossett98:q-carry-save}.

\begin{figure}
\centerline{\hbox{
\includegraphics[width=8.6cm]{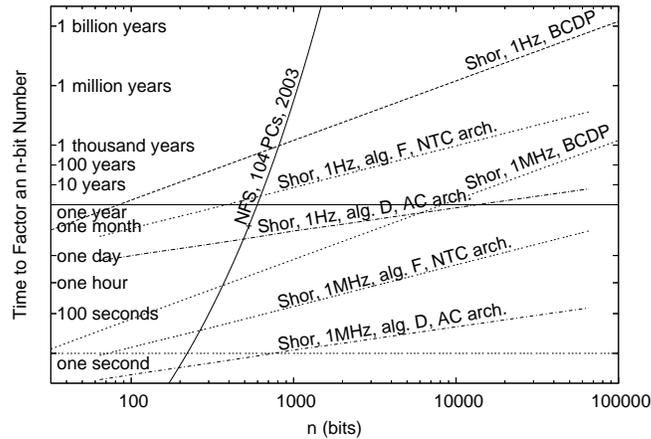}}}
\caption{Scaling of number field sieve (NFS) and Shor's algorithms for
  factoring, using faster modular exponentiation
  algorithms.}
\label{fig:scaling-fast}
\end{figure}

We have composed several algorithm variants, {\bf A} through {\bf F},
as well as investigated concurrent and parallel versions of the
original Vedral-Barenco-Ekert (VBE)~\cite{vedral:quant-arith} and BCDP
algorithms~\cite{van-meter04:fast-modexp}; only the fastest for our AC
and NTC architectures are presented here.  Four parameters control the
behavior of the algorithm variants and how well they match a
particular architecture.  These parameters include the choice of type
of adder and the amount of space required.  Algorithm variant {\bf D}
is tuned for AC using the conditional-sum adder, and {\bf F} is tuned
for NTC using the Cuccaro-Draper-Kutin-Moulton (CDKM) carry-ripple
adder~\cite{cuccaro04:new-quant-ripple}.  We have optimized the
parameter settings for each individual data point, though the
differences are just barely visible on our log-log plot.  The values
reported here for both algorithms are calculated using $2n^2$ qubits
of storage to exponentiate an $n$-bit number, the largest number of
qubits our algorithms can effectively use.  The primary
characteristics of the algorithms shown in
Figure~\ref{fig:scaling-fast} are summarized in
Table~\ref{tab:algorithms}.  The table lists the number of
multiplication units executing concurrently, the space, measured in
number of logical qubits, the concurrency, or number of logical
operations taking place at the same time, and the overall circuit
depth, or time, measured in gates.
\begin{table*}
\tbl{Composition of our algorithms.
\label{tab:algorithms}}
{\begin{tabular}{@{}lccccc@{}} \hline
algorithm & adder &  multipliers ($s$) & space & concurrency & depth \\ \hline
conc. BCDP & BCDP & 1 & $5n+3$ & 2 & $\sim 54n^3$ \\
algorithm {\bf D}
& cond. sum & $\sim n/4$  & $2n^2$ & $\sim n^2$ &	$\sim 9n\log_2^2(n)$ \\
algorithm {\bf F}
& CDKM & $\sim n/4$ & $2n^2$ & $\sim 3n/4$ & $\sim 20n^2\log_2(n)$	\\
\hline
\end{tabular}}
\end{table*}

Figure~\ref{fig:scaling-fast} shows our results for our faster
algorithms.  We have kept the 1~Hz and 1~MHz lines for BCDP, and added
matching lines for our fastest algorithms on the AC and NTC
architectures.  For AC, our algorithm {\bf D} requires a clock rate of
only about 0.3~Hz to factor the same 576-bit number in one month.  For
NTC, using our algorithm {\bf F}, a clock rate of around 27~Hz is
necessary.  The graph shows that, for problem sizes larger than 6,000
bits, our algorithm {\bf D} is one million times faster than the basic
BCDP algorithm, and algorithm {\bf F} is one thousand times faster.
For very large $n$, the latency of {\bf D} is $\sim 9n\log_2^2(n)$.
The latency of {\bf F} is $\sim 20n^2\log_2(n)$.

This relationship of architecture and algorithm has obvious
architectural implications: concurrency is critical, and support for
long-distance gates is important.

\secton{Discussion}

A fast clock speed is obviously also important for a fast algorithm;
however, it remains an open question whether those quantum computing
technologies which feature naturally fast \emph{physical} quantum
gates will have the fastest overall algorithm speed.  All quantum
computing technologies feature some level of decoherence, requiring
resources for quantum error
correction~\cite{steane02:ft-qec-overhead,devitt04:_shor_qec_simul,nielsen-chuang:qci}. As
an example, quantum computers based on Josephson junctions are likely
to have extremely fast single-qubit and two-qubit gates, with a
physical clock rate at the gigahertz level, as demonstrated in recent
experiments~\cite{nakamura}. However, the single-qubit decoherence
time is only about 1~$\mu$s for the most coherent superconducting
qubits~\cite{vion}. Although ``fast,'' the difficulty in long-term
qubit storage and the needed resources for fault tolerant operation
may be quite large, so these implementations might make excellent
processors with poor memories. In sharp contrast, NMR-based
approaches~\cite{kane:nature-si-qc,ladd:si-nmr-qc} are quite slow,
with nuclear-nuclear interactions in the kilohertz range. However, the
much longer coherence times of nuclei~\cite{ladd:coherence-time} make
the use of NMR-based qubits as memory substantially
easier~\cite{itoh:all-si}. Ion trap implementations have the benefit
of faster single-qubit-gate, two-qubit-gate, and qubit-measurement
speeds with longer coherence times, but the added complication of
moving ionic qubits from trap to trap
physically~\cite{kielpinski:large-scale} or exchanging their values
optically~\cite{ionphoton} complicates the picture for the
application-level clock rate.  New physical proposals for overcoming
speed and scalability obstacles continue to be developed, leaving the
ultimate hardware limitations on clock speed and its relation to
algorithm execution time uncertain.

\secton{Conclusions}

We have shown that the actual execution time of Shor's algorithm is
dependent on the important features of concurrent gate execution,
available number of qubits, interconnect topology, and clock speed, as
well as the critical choice of an architecture-appropriate arithmetic
algorithm.  Our algorithms have shown a speed-up factor ranging from
nearly 13,000 for factoring a 576-bit number to one million for a
6,000-bit number.

\section*{Acknowledgments}

The authors wish to thank Eisuke Abe, Kevin Binkley, Fumiko Yamaguchi,
Seth Lloyd, Kae Nemoto, and W. J. Munro for helpful discussions.


\begin{thebibliography}{10}

\bibitem{nielsen-chuang:qci}
M.~A. Nielsen and I.~L. Chuang, {\em Quantum Computation and Quantum
  Information} (Cambridge University Press, Cambridge, 2000).

\bibitem{divincenzo95:qc}
D.~P. DiVincenzo, Science {\bf 270},  255  (1995).

\bibitem{van-meter:qarch-impli}
R.~Van{ }Meter and M.~Oskin.
\newblock {\em J. Emerging Tech. in Comp. Sys.}, 2(1), Jan. 2006.

\bibitem{shor:factor}
P.~W. Shor,  in {\em Proc. 35th Symposium on Foundations of Computer Science}
  (IEEE Computer Society Press, Los Alamitos, CA, 1994), pp.\ 124--134.

\bibitem{knuth:v2-3rd}
D.~E. Knuth, {\em The Art of Computer Programming, volume 2 / Seminumerical
  Algorithms}, 3rd ed. (Addison-Wesley, Reading, MA, 1998).

\bibitem{beckman96:eff-net-quant-fact}
D. Beckman, A.~N. Chari, S. Devabhaktuni, and J. Preskill, Phys. Rev. A {\bf
  54},  1034  (1996).

\bibitem{rsa04-factoring}
RSA{ }Security{ }Inc., web page, 2004,
  http://www.rsasecurity.com/rsalabs/node.asp?id=2096.

\bibitem{kunihiro05:_exact_ieice}
N. Kunihiro,
IEICE Trans. Fundamentals, \textbf{E88-A}(1):105--111, (2005).

\bibitem{fowler04:_shor_implem}
A.~G. Fowler, S.~J. Devitt, and L.~C. Hollenberg, Quantum Information and
  Computation {\bf 4},  237  (2004).

\bibitem{loss:qdot-comp}
D. Loss and D.~P. DiVincenzo, Phys. Rev. A {\bf 57},  120  (1998).

\bibitem{kane:nature-si-qc}
B.~E. Kane, Nature {\bf 393},  133  (1998).

\bibitem{ladd:si-nmr-qc}
T.~D. Ladd {\it et~al.}, Phys. Rev. Lett. {\bf 89},  17901  (2002).

\bibitem{vedral:quant-arith}
V. Vedral, A. Barenco, and A. Ekert, Phys. Rev. A {\bf 54},  147  (1996).

\bibitem{ercegovac-lang:dig-arith}
M.~D. Ercegovac and T. Lang, {\em Digital Arithmetic} (Morgan Kaufmann, San
  Francisco, CA, 2004).

\bibitem{van-meter04:fast-modexp}
R. Van{ }Meter and K.~M. Itoh, Phys. Rev. A {\bf 71},  052320  (2005).

\bibitem{draper04:quant-carry-lookahead}
T.~G. Draper, S.~A. Kutin, E.~M. Rains, and K.~M. Svore, A Logarithmic-Depth
  Quantum Carry-Lookahead Adder, http://arXiv.org/quant-ph/0406142 (2004).

\bibitem{gossett98:q-carry-save}
P. Gossett, Quantum Carry-Save Arithmetic, http://arXiv.org/quant-ph/9808061
  (1998).

\bibitem{cuccaro04:new-quant-ripple}
S.~A. Cuccaro, T.~G. Draper, S.~A. Kutin, and D.~P. Moulton, A new quantum
  ripple-carry addition circuit, http://arXiv.org/quant-ph/0410184, 2004.

\bibitem{steane02:ft-qec-overhead}
A.~M. Steane, Phys. Rev. A {\bf 68},  042322  (2003).

\bibitem{devitt04:_shor_qec_simul}
S.~J. Devitt, A.~G. Fowler, and L.~C. Hollenberg, Simulations of {Shor's}
  algorithm with implications to scaling and quantum error correction,
  http://arXiv.org/quant-ph/0408081, 2004.

\bibitem{nakamura} T. Yamamoto, \textit{et al.}, Nature
\textbf{425}, 941 (2003).

\bibitem{vion} D. Vion, \textit{et al.}, Science \textbf{296}, 886
(2002).

\bibitem{ladd:coherence-time}
T.~D. Ladd {\it et~al.}, Phys. Rev. B {\bf 71}, 014401 (2005).

\bibitem{itoh:all-si}
K.~M. Itoh, Solid State Comm. {\bf 133}, 747 (2005).

\bibitem{kielpinski:large-scale}
D. Kielpinski, C. Monroe, and D.~J. Wineland, Nature {\bf 417},  709  (2002).

\bibitem{ionphoton}{B.B. Blinov, D. L. Moehring, L.-M. Duan, and
C. Monroe, Nature \textbf{428}, 153 (2004).}

\bibitem{duan}L. M. Duan, Phys. Rev. Lett. \textbf{93}, 100502 (2004).

\end{thebibliography}

\end{document}